\documentclass[final,5p,times,twocolumn]{elsarticle}

\journal{Physics Letters B}

\usepackage{amsmath,amssymb}
\usepackage{bm,comment,braket}
\usepackage[x11names]{xcolor}
\usepackage[T1]{fontenc}
\usepackage{colortbl}
\usepackage{ulem}
\usepackage{cancel}
\usepackage{hyperref}

\allowdisplaybreaks

\newcommand{\del}{\partial}
\newcommand{\beq}{\begin{eqnarray}}
\newcommand{\eeq}{\end{eqnarray}}

\renewcommand\sout{\bgroup \color{red} \ULdepth=-.5ex \ULset}

\newcommand{\p}{\partial}

\begin{document}

\begin{frontmatter}

\title{
Revisiting 
the Wess-Zumino-Witten Term in Nuclear and Quark Matter 
\\ 
under Magnetic Fields and Rotation
}

\author[a,b]{Markus~A.~G.~Amano}
\ead{markus(at)oyama-ct.ac.jp}

\author[b,c,d]{Minoru~Eto}
\ead{meto(at)sci.kj.yamagata-u.ac.jp}

\author[e,c,d]{Muneto~Nitta}
\ead{mune.nitta(at)gmail.com}

\author[f]{Shin~Sasaki}
\ead{shin-s(at)kitasato-u.ac.jp}

\affiliation[a]{
organization={Department of General Education, National Institute of Technology, Oyama College},
addressline={Ooazanakakuki 777},
city={Oyama City},
postcode={323-0806},
state={Tochigi},
country={Japan}
}

\affiliation[b]{
organization={Department of Physics, Yamagata University},
addressline={Kojirakawa-machi 1-4-12},
city={Yamagata},
postcode={990-8560},
state={Yamagata},
country={Japan}
}

\affiliation[c]{
organization={Research and Education Center for Natural Sciences, Keio University},
addressline={4-1-1 Hiyoshi},
city={Yokohama},
postcode={223-8521},
state={Kanagawa},
country={Japan}
}

\affiliation[d]{
organization={International Institute for Sustainability with Knotted Chiral Meta Matter (WPI-SKCM$^2$), Hiroshima University},
addressline={1-3-2 Kagamiyama},
city={Higashi-Hiroshima},
postcode={739-8531},
state={Hiroshima},
country={Japan}
}

\affiliation[e]{
organization={Department of Physics, Keio University},
addressline={4-1-1 Hiyoshi},
city={Yokohama},
postcode={223-8521},
state={Kanagawa},
country={Japan}
}

\affiliation[f]{
organization={Department of Physics, Kitasato University},
city={Sagamihara},
postcode={252-0373},
country={Japan}
}

\begin{abstract}
We study anomalous Wess-Zumino-Witten  terms associated with the chiral anomaly in dense QCD matter under magnetic fields and rotation. 
By introducing electromagnetic, baryon number, and isospin background gauge fields, we write down the topological couplings of neutral mesons for the $N_f=2$ and $N_f=3$ cases.
The resulting terms contain characteristic contributions proportional to 
$\vec{B} \cdot \vec{\nabla} \phi$
and 
$\vec{\Omega} \cdot \vec{\nabla} \phi$, 
where $\phi$ denotes $\pi^0$, $\eta$, or $\eta'$. These results are relevant to chiral soliton lattices in dense rotating matter.
\end{abstract}


\end{frontmatter}

\section{Introduction} \label{sec:introduction}

Quantum anomalies play a profound role in quantum field theory, governing a variety of nonperturbative phenomena that cannot be inferred from classical symmetries alone. In low-energy effective theories of Nambu–Goldstone bosons, anomalies are encoded in the Wess–Zumino–Witten (WZW) term \cite{Wess:1971yu,Witten:1983tx}, which reproduces the anomalous Ward identities of the underlying microscopic theory. Since its formulation, the WZW term has provided a unified framework for understanding anomaly-induced processes in particle, nuclear, and condensed-matter physics.

At finite density, the physical significance of anomalies becomes even more striking. In the presence of external magnetic fields or rotation, the WZW term generates couplings between Nambu–Goldstone fields and background gauge fields that are absent in the ordinary derivative expansion \cite{Son:2004tq}. These couplings give rise to anomalous currents \cite{Goldstone:1981kk} and induce nontrivial topological structures in the ground state \cite{Son:2007ny}. Well-known examples include the chiral magnetic effect, the chiral vortical effect, and anomaly-induced baryon number transport. More generally, anomalies can qualitatively alter the phase structure of matter by stabilizing spatially modulated or topologically nontrivial configurations \cite{Son:2007ny}.

Recent studies have revealed that anomaly-induced terms can determine the true
ground states of strongly interacting matter under extreme conditions. In dense
QCD subjected to strong magnetic fields, the WZW term favors inhomogeneous pion
configurations such as the chiral soliton lattice (CSL)
\cite{Son:2007ny,Eto:2012qd,Brauner:2016pko}, whose existence is directly tied
to the interplay between baryon density and the axial anomaly. Various aspects
of CSLs in field theory were studied, such as thermal
fluctuations \cite{Brauner:2017uiu, Brauner:2017mui,
Brauner:2021sci, Brauner:2023ort}, quantum and thermal nucleation
\cite{Eto:2022lhu,Higaki:2022gnw,Eto:2025ebz}, an incommensurate crystal of the
pion and $\eta$ meson \cite{Qiu:2023guy}, and also CSLs in QCD-like theories
\cite{Brauner:2019rjg,Brauner:2019aid},
supersymmetric QCD \cite{Nitta:2024xcu}, and
holographic QCD \cite{Amano:2025iwi}. Similar mechanisms can stabilize
Skyrmionic textures \cite{Amari:2025twm}, domain-wall
Skyrmions
\cite{Eto:2023lyo,Eto:2023wul,Amari:2024fbo,Amari:2024mip,Copinger:2025rpo},
baryonic crystals~\cite{Evans:2022hwr,Evans:2023hms},
vortex-Skyrmions~\cite{Qiu:2024zpg,Hamada:2025inf,Hamada:2026uec,Mameda:2026kyp},
and other topological solitons. 
A common outstanding feature of these ground states is that they are stabilized
in the presence of the anomaly and that they carry baryon
number. Furthermore, in rotating systems, anomaly-induced couplings generate
analogous effects through vorticity, leading to novel phases that would
otherwise be absent
\cite{Huang:2017pqe,Nishimura:2020odq,Chen:2021aiq,Eto:2021gyy,Eto:2023tuu,Eto:2023rzd,Evans:2025jsa}.

These developments suggest a broader perspective: rather than merely producing anomalous currents on top of a given background, anomalies can themselves dictate the structure of the vacuum or ground state. In this sense, the WZW term acts as a driving force for the emergence of topological matter at finite density, magnetic field, and rotation.

In this work, we derive the anomaly-induced WZW terms for dense QCD matter in the presence of external magnetic fields %$\mathbf{B}$ 
$\vec{B}$ and rotation 
%$\boldsymbol{\Omega}$
$\vec{\Omega}$. 
By introducing electromagnetic, baryon-number, and isospin background gauge fields, we obtain the topological couplings of neutral mesons for both the $N_f=2$, $N_f=3$ 
and also the general $N_f$ cases. The resulting anomalous terms contain characteristic contributions proportional to 
$\vec{B} \cdot \vec{\nabla} \phi$
and 
$\vec{\Omega} \cdot \vec{\nabla} \phi$
, where $\phi$ denotes neutral Nambu–Goldstone fields such as $\pi^0$, $\eta$, and $\eta'$. These couplings provide the anomaly-induced driving terms for spatially modulated configurations and are particularly relevant to the formation of CSLs and related topological ground states in dense rotating matter.

\if0
In this work, we revisit anomaly-induced WZW terms in dense QCD matter under external magnetic fields and rotation. By introducing electromagnetic, baryon-number, and isospin background gauge fields, we systematically derive the anomalous couplings of neutral mesons for both the $N_f=2$ and $N_f=3$ cases. The resulting effective Lagrangians contain characteristic terms proportional to $ \mathbf{B}\cdot\nabla\phi$ and $ \boldsymbol{\Omega}\cdot\nabla\phi$, where $\phi$ denotes the neutral meson fields $\pi^0$, $\eta$, and $\eta'$. 
%We also discuss the role of nontrivial Weyl-group orbits and present a complementary derivation based on ($U(1)$) anomalies. 
Our results provide a unified description of anomaly-induced topological couplings in dense matter and offer a useful framework for studying chiral soliton lattices and other inhomogeneous phases in magnetic and rotating environments.
\fi
\if0
In this work, we investigate the role of anomaly-induced WZW terms in determining the ground states of finite-density systems in the presence of external magnetic fields and rotation. We develop a unified framework in which the competition between conventional energy contributions and anomaly-induced topological terms can be analyzed systematically. 
Anomalies do not merely modify existing phases; they can protect entirely new solitonic ground states.
%This allows us to identify the conditions under which anomalous effects dominate and stabilize nontrivial solitonic ground states.
\fi

The remainder of this paper is organized as follows. In Sec.~\ref{sec:WZW_term}, we derive the WZW terms under magnetic fields and rotation. In Sec.~\ref{sec:anomaly}, we revisit the same couplings from the viewpoint of $\mathrm{U}(1)$ anomalies and discuss the resulting anomalous currents.
Section \ref{sec:conclusion} is devoted to a summary and discussion.
In \ref{sec:rotation}, we discuss the leading order correspondence between rotation and an effective background baryon gauge field.

\section{Wess-Zumino-Witten term} \label{sec:WZW_term}
We consider QCD with $N_c$ colors and $N_f$ massive quarks in background gauge fields.
The quark mass term breaks $\mathrm{U}(N_f)_L \times \mathrm{U}(N_f)_R$ flavor symmetry down to the diagonal subgroup $\mathrm{U}(N_f)$.

\subsection{Backgrounds in WZW term}
The relevant part of the WZW term is given by
\begin{align}
S_{\text{WZW}} = S_{\text{WZW}}^{(0)} - \frac{N_c}{48 \pi^2} \int \! Z,
\end{align}
where $S_{\text{WZW}}^{(0)}$ is the term that does not depend on background gauge fields, and 
$Z$ in our convention is given by
\begin{align}
Z =& \ 
i \,
\mathrm{Tr}
\Bigg[
\Big(
A_R d A_R + d A_R A_R - i A_R^3
\Big)
\Big(
-i \, 
\Sigma^{\dagger} A_L \Sigma + \Sigma^{\dagger} d \Sigma
\Big) 
- \text{p.c.}
%\tag{A} \label{A} 
\notag 
\\
& \qquad
- d A_R \, d \Sigma^{\dagger} A_L \Sigma 
%- \text{p.c.}
%\tag{B} \label{B} \\
%& \qquad
+ i \, A_R (d \Sigma^{\dagger} \, \Sigma)^3 
%- \text{p.c.}
%\tag{C} \label{C} \\
%& \qquad
+ \frac{1}{2} (A_R d \Sigma^{\dagger} \, \Sigma)^2 
- \text{p.c.}
%\tag{D} \label{D} 
\notag \\
& \qquad
+ \Sigma A_R \Sigma^{\dagger} A_L d \Sigma \, d \Sigma^{\dagger} 
%- \text{p.c.}
%\tag{E} \label{E} 
%\notag \\
%& \qquad
+ i \, A_R d \Sigma^{\dagger} \, \Sigma A_R \Sigma^{\dagger} A_L \Sigma - \text{p.c.}
%\tag{F} \label{F} \\
\notag \\
& \qquad
- \frac{1}{2} (A_R \Sigma^{\dagger} A_L \Sigma)^2
\Bigg]
%\tag{G} \label{G},
\end{align}
where $\Sigma = e^{i \phi/f}$ is the meson field and 
$A_L, A_R$ are gauge fields of $\mathrm{U}(N_f)_L$ and $\mathrm{U}(N_f)_R$, respectively.
The symbol p.c. stands for terms with $A_L \leftrightarrow A_R$, $\Sigma
\leftrightarrow \Sigma^{\dagger}$ \cite{Harvey:2007ca}.
We note that this is also derived from the Chern–Simons term of $D8$-branes in a holographic QCD model \cite{Sakai:2004cn}.

We introduce backgrounds of the electromagnetic gauge field $A_Q$ together with the isospin and 
the baryon number gauge fields $A_I, A_B$:
%\begin{align}
%A_L = A_R =  
%A_{Q} \, Q
%+
%A_{I} \, Q_I
%+
%A_B \, Q_B
%\label{eq:background_gauge_fields}
%\end{align}
\begin{align}
A_L = A_R = 
e Q \, A_{Q} 
-
Q^I \, A_{I}
-
Q^B \, A_B.
\label{eq:background_gauge_fields}
\end{align}

Here $e Q$, $Q_I$ and $Q_B = \frac{1}{N_c} \mathbf{1}$ are the electromagnetic, isospin, and baryon charge matrices, respectively.
We have assumed that they are in the Cartan part of $\text{U}(N_f)$.
%$\tilde{A}_B = \frac{A_B}{N_c}$.
Then $Z$ becomes
\begin{align}
Z = Z_1 + Z_2
\label{eq:Z_term}
\end{align}
where 
\begin{align}
Z_1 =& \ 
- 
\mathrm{Tr}
\Big[
\Big(
e Q \, A_Q - Q^I \, A_I
\Big) (R^3 + L^3)
\Big]
\notag \\
& \
- 2 i \, \mathrm{Tr}
\Bigg[
\Big(
e Q \, A_Q - Q^I \, A_I
\Big)
\Big(
Q \, d A_Q - Q^I \, d A_I
\Big)
\Big(
R + L
\Big)
\Bigg]
\notag \\
& \ 
- i \mathrm{Tr}
\Bigg[
\Big(
e Q \, d A_Q - Q^I \, d A_I
\Big)
d \Sigma^{\dagger}
\Big(
e Q \, A_Q - Q^I \, A_I
\Big)
\Sigma
\notag \\
& \qquad \quad
-
\Big(
e Q \, d A_Q - Q^I \, d A_I
\Big)
d \Sigma 
\Big(
e Q \, A_Q - Q^I \, A_I
\Big)
\Sigma^{\dagger}
\Bigg]
\notag \\
& \ 
- 2 \, A_B \, 
\Bigg\{
\mathrm{Tr} [- Q_B \, R^3]
+ 3 i d \,
\mathrm{Tr}
\Big[
- Q_B \,
\Big(
e Q \, A_Q - Q^I \, A_I
\Big) 
(R + L)
\Big]
\Bigg\}
\notag \\
& \ 
- 3 i \, \mathrm{Tr}
\Bigg[
A_B \, d A_B \, (Q^B)^2 (R + L)
\Bigg]
- \mathrm{Tr} 
\Bigg[
e Q \, A_Q
\Big(
R Q^I A_I R
- L Q^I A_I L
\Big)
\Bigg]
\notag \\
& \
+ 3 i \, d\, \mathrm{Tr}
\Bigg[
\Big(
e Q \, A_Q - Q^I \, A_I
\Big)
(-Q^B) \, A_B \, (R + L)
\Bigg]
\label{eq:Z1}
\end{align}
%%%%%%%%%%%%%%
and
%%%%%%%%%%%%%%
\begin{align}
Z_2 =& \
2 i \, \mathrm{Tr}
\Bigg[
-
e^2 Q^2 \, A_Q d A_Q 
\Big(
\Sigma^{\dagger} Q^I A_I \Sigma - \Sigma Q^I A_I  \Sigma^{\dagger}
\Big)
\Bigg]
\notag \\
& \
+ 2 \, 
\mathrm{Tr}
\Bigg[
e Q \, A_Q 
\Big(
\Sigma e Q d A_Q Q^I A_I \Sigma^{\dagger}
- \Sigma^{\dagger} e Q d A_Q Q^I A_I \Sigma
\notag \\
& \qquad \qquad \qquad \qquad 
- \Sigma (Q^I)^2 d A_I A_I \Sigma^{\dagger}
+ \Sigma^{\dagger} (Q^I)^2 d A_I A_I \Sigma
\Big)
\Bigg]
\notag \\
& \ 
+ 2 \, \mathrm{Tr}
\Bigg[
Q^B A_B Q^I d A_I 
\Big(
- \Sigma^{\dagger} Q^I A_I \Sigma
+
\Sigma Q^I A_I \Sigma^{\dagger}
\notag \\
& \qquad \qquad \qquad \qquad \qquad
+
\Sigma^{\dagger} e Q A_Q \Sigma
-
\Sigma e Q A_Q\Sigma^{\dagger}
\Big)
\Bigg]
\notag \\
& \ 
+ 2 \, \mathrm{Tr}
\Bigg[
e Q A_Q Q^I d A_I
\Big(
\Sigma^{\dagger} Q^I A_I \Sigma - \Sigma Q^I A_I  \Sigma^{\dagger}
\Big)
\Bigg]
\notag \\
& \ 
+ 4 \, \mathrm{Tr}
\Bigg[
e Q A_Q Q^B d A_B
\Big(
\Sigma^{\dagger} Q^I A_I \Sigma
-
\Sigma Q^I A_I \Sigma^{\dagger}
\Big)
\Bigg]
\notag \\
& \ 
- i \,
\mathrm{Tr} 
\Bigg[
e Q A_Q 
(R Q^I A_I R
-
L Q^I A_I L)
\Bigg]
\notag \\
& \ 
- i \, 
\mathrm{Tr}
\Bigg[
\Big(
\Sigma^{\dagger} Q^I A_I \Sigma e Q A_Q
-
e Q A_Q \Sigma^{\dagger} Q^I A_I \Sigma
\Big) 
R^2
\notag \\
& \qquad \qquad 
-
\Big(
\Sigma Q^I A_I \Sigma^{\dagger} e Q A_Q
-
e Q A_Q \Sigma Q^I A_I \Sigma^{\dagger}
\Big) L^2
\Bigg]
\notag \\
& \ 
+ 
\mathrm{Tr}
\Bigg[
\Big(
e Q A_Q - Q^I A_I - Q^B A_B 
\Big) 
+ 
\Big(
e Q A_Q - Q^I A_I - Q^B A_B 
\Big) \Sigma
\notag \\
& \qquad \qquad \times 
\Big(
e Q A_Q - Q^I A_I - Q^B A_B
\Big) \Sigma^{\dagger}
- (\Sigma \leftrightarrow \Sigma^{\dagger})
\Bigg].
\label{eq:Z2}
\end{align}
Here we have defined $L = \Sigma \, d \Sigma^{\dagger}$ and $R = d \Sigma^{\dagger} \Sigma$.

\subsection{$N_f = 2$ case}
Now we introduce $\pi^0$ and $\eta'$ in the $N_f=2$ WZW term:
\begin{align}
\Sigma 
= e^{i \left(\frac{\pi^0}{f_{\pi}} \tau^3 + \frac{\eta'}{f_{\eta'}} \mathbf{1}_2 \right)}.
\end{align}
In the following, 
we use the mostly minus convention for the metric $\eta_{\mu \nu} = \mathrm{diag} (1,-1,-1,-1)$ and 
assume that $\del_0 \Sigma = 0$.

We introduce the background gauge fields \eqref{eq:background_gauge_fields} with the charge matrices:
\begin{align}
Q := \frac{\tau^3}{2} + \frac{1}{6}, \qquad Q_I := \frac{\tau^3}{2}.
\end{align}
The isospin and the baryon chemical potentials are introduced as $A_{I} = (\mu_I, 0,0,0)$ and $A_B = (\mu_B,0,0,0)$,
where $\mu_I$ and $\mu_B$ are constants.
The effect of rotation on them is incorporated by a Lorentz boost by $\vec{v} = \vec{\Omega} \times \vec{x}$, where $\vec{\Omega}$ is a constant angular velocity \cite{Son:2004tq, Abramchuk:2018jhd}.
Assuming $|v| = |\vec{x}| |\vec{\Omega}| \ll c = 1$, we have
$
A_{I}^{\mu} = \mu_I (1, \vec{\Omega} \times \vec{x})
$ and $
A_{B}^{\mu} = \mu_B (1, \vec{\Omega} \times \vec{x})
$ in the laboratory frame.
In \ref{sec:rotation}, we show that 
up to linear order in $\Omega$, these gauge fields can be interpreted as being induced in a rotating frame.
Throughout this work, we focus on the slow rotation regime 
$|\vec{\Omega} \times \vec{x}| \ll 1$ and keep only terms linear in $\Omega$. 
The correspondence between rotation and an effective background magnetic gauge field should therefore be understood only at this order.
We further introduce a constant background magnetic field $\vec{B}$ in either the rest or the rotating frame under consideration.
In components, they are given by
\begin{align}
A_0 =& \ 
-(A_I)_0 \frac{\tau^3}{2}
- \frac{(A_B)_0}{N_c} \mathbf{1}
= 
- \mu_I \frac{\tau^3}{2} 
- \frac{\mu_B}{N_c} \mathbf{1},
%\label{eq:nf2_a0_components} 
\notag \\
A_i =& \ 
e Q (A_Q)_i - (A_I)_i \frac{\tau^3}{2} - \frac{(A_B)_i}{N_c} \mathbf{1}
\notag \\
=& \ 
\frac{1}{2} \varepsilon_{ijk} B_j x^k Q
-
\mu_I v_i \frac{\tau^3}{2}
-
\frac{\mu_B}{N_c} v_i \mathbf{1},
\notag \\
F_{i0} =& \ 0,
\notag \\
F_{ij} =& \ 
e Q (F_{Q})_{ij}
-
(F_{I})_{ij} \frac{\tau^3}{2}
-
\frac{1}{N_c} (F_B)_{ij} \mathbf{1}
\notag \\
=& \ 
e Q \varepsilon_{ijk} B^k
+ \mu_I \varepsilon_{ijk} \Omega^k \tau^3
+ \frac{2}{N_c} \mu_B \varepsilon_{ijk} \Omega^k \mathbf{1}.
\label{eq:Nf2_backgrounds}
\end{align}
Substituting Eq.~\eqref{eq:Nf2_backgrounds} into Eq.~\eqref{eq:Z2} results in a vanishing contribution, whereas 
the nonzero terms come from Eq.~\eqref{eq:Z1}.
We find
%\MA{The factor on $\Omega$ is of by a factor of $1/2$ where $\Omega\rightarrow \Omega/2$ would fix it.}
\begin{align}
\mathcal{L}_{\text{WZW}} &= 
\frac{e}{4 \pi^2 f_{\pi}}  
\Big(
\mu_B + 
 \frac{N_c}{6} \mu_I
\Big) \vec{B} \cdot \vec{\nabla} \pi^0 
+ \frac{\mu_B \mu_I}{2 \pi^2 f_{\pi}} 
\vec{\Omega} \cdot \vec{\nabla} \pi^0
\notag \\
+
&
\frac{e}{4 \pi^2 f_{\eta'}}  
\Big(
\frac{\mu_B}{3} +  \frac{N_c}{2} \mu_I
\Big) \vec{B} \cdot \vec{\nabla} \eta'
+
\frac{1}{4 \pi^2 f_{\eta'}}  
\Big(
\frac{2 \mu_B^2}{N_c} 
+
\frac{N_c \mu_I^2}{2}
\Big) 
\vec{\Omega} \cdot \vec{\nabla} \eta'.
\label{eq:WZW_Nf=2}
\end{align}
We have performed integration by parts 
so that $A_Q$ implicitly appears via $F_Q$ only.
The result is consistent with the one in \cite{Son:2004tq}.

\subsection{$N_f=3$ case}
We next examine the $N_f = 3$ case where the neutral mesons are introduced as 
\begin{align}
\Sigma =
e^{
i 
\Big(
\frac{\eta'}{f_{\eta'}} \lambda^0 + \frac{\pi^0}{f_{\pi}} \lambda^3 + \frac{\eta}{f_{\eta}} \lambda^8
\Big)
}.
\end{align}
Here $\lambda^{\alpha} \, (\alpha = 0,3,8)$ are the Cartan part of the Gell-Mann matrices:
\begin{align}
\lambda^0 = \sqrt{\frac{2}{3}} \mathbf{1}_3,
\
\lambda^3 = 
\left(
\begin{array}{ccc}
1 & & \\
& - 1 & \\
& & 0
\end{array}
\right),
\
\lambda^8 =
\frac{1}{\sqrt{3}}
\left(
\begin{array}{ccc}
1 & & \\
& 1 & \\
& & -2
\end{array}
\right).
\end{align}
We introduce the background fields \eqref{eq:background_gauge_fields} and \eqref{eq:Nf2_backgrounds}, where the charge matrices are 
\begin{align}
Q := \frac{1}{2} \lambda^3 + \frac{1}{2 \sqrt{3}} \lambda^8,
\qquad
Q_I := \frac{1}{2} \lambda^3.
\end{align}
We again obtain vanishing contributions from $Z_2$, and $Z_1$ gives the following topological terms:
\begin{align}
\mathcal{L} =&\
 \frac{e}{4\pi^2 f_{\pi}}
\left(
\mu_B  + \frac{N_c \mu_I}{6}
\right) 
\vec{B} \cdot \vec{\nabla} \pi^0
+ \frac{\mu_B \mu_I}{2\pi^2 f_{\pi}} 
\vec{\Omega} \cdot \vec{\nabla} \pi^0
\notag\\
&\
+ \frac{e}{4\pi^2 f_{\eta}}
\left(
\frac{\mu_B}{\sqrt{3}} + \frac{N_c \mu_I}{2\sqrt{3}}
\right) 
\vec{B} \cdot \vec{\nabla} \eta
+ \frac{\sqrt{3}\, N_c \mu_I^2}{24\pi^2 f_{\eta}} 
\vec{\Omega} \cdot \vec{\nabla} \eta
\notag\\
&\
+ \frac{\sqrt{6} e}{24\pi^2 f_{\eta'}} N_c \mu_I\, B_k \partial_k \eta'
+ \frac{\sqrt{6}}{24\pi^2 f_{\eta'}}
\left(
\frac{6\mu_B^2}{N_c} + N_c \mu_I^2
\right) 
\vec{\Omega} \cdot \vec{\nabla} \eta' .
\end{align}

\subsection{General $N_f$ case}
Given the results for the cases of $N_f = 2$ and $N_f = 3$ presented above, 
we derive the topological term for the general case of $N_f$.
We introduce background gauge fields of the form:
\begin{align}
A_L = A_R = A = \mu \, dt + \frac{1}{2} \varepsilon_{ijk} V_j x^k dx^i,
\end{align}
where $\mu = \sum_{A} \mu_A Q_A$ and 
%$V_i = B_i Q + 2 \mu_I \Omega_i Q_I + 2 \mu_B \Omega_i Q_B$ 
$V_i = e B_i Q 
-
2 \mu \Omega_i$
are in the Cartan subalgebra of $\mathfrak{u}(N_f)$.
Here, $Q$ is the electric charge matrix and $Q_A$ are the Cartan generators.
We also assume that the neutral meson fields $\phi$ in the Cartan part are static.
Then, the WZW Lagrangian is evaluated as 
\begin{equation}
\mathcal L_{\text{WZW}} =  \frac{N_c}{8\pi^2} \mathrm{Tr} \left( \mu \vec V' \cdot \vec\nabla\phi \right),
\end{equation}
where 
$V'_i = 2 (B_i Q - \mu \Omega_i)$.
Due to the partial application of integration by parts, the resulting $V$ effectively becomes $V'$ with the $B_i$ term scaled by a factor of 2.

%%%%%%%%%%%%%%%
\section{WZW-like terms in low energy effective theories}\label{sec:anomaly}

Here, we reconsider axial anomalies from an elementary point of view.
We will see that the results in the previous section are correctly reproduced.
An advantage of the approach presented in this section is that we can treat the WZW-like terms not only for the hadron phase at low density but also for the color superconducting phase at high density.

\subsection{Some basics}

\subsubsection{Axial anomalies}

The part of the QCD Lagrangian with $N_f$ quarks coupled to background $\text{U}(1)_V$ gauge fields is given by\,\footnote{In this section, we essentially follow the notation of Ref.~\cite{Son:2004tq}.}
\begin{eqnarray}
    {\cal L}_V = \sum_{i=1}^{N_f} \bar q_i \gamma^\mu 
    \left(i \p_\mu - A_{i\,\mu}\right) q_i\,.
\end{eqnarray}
The $\text{U}(1)_{i\,V}$ transformation shifts the phase of the $i$-th flavor as 
\begin{eqnarray}
    q_i \to e^{i\beta_i} q_i\,,\quad A_i \to A_{i\,\mu} - \partial_\mu\beta_i\,.
\end{eqnarray}
These transformations are not anomalous, because they are vector-like symmetries.
The only physical gauge field is the one corresponding to $\text{U}(1)_{\text{EM}}$, and all the rest are fictitious (spurion) gauge fields.
For example, the spurion $\text{U}(1)_B$ gauge field $A_{B\mu}$ is introduced by
\begin{eqnarray}
    A_\mu = -Q^B A_{B\mu}\,,\quad
    Q^B = \frac{1}{N_C}{\bf 1}_{N_f}\,.
\end{eqnarray}
Then, the quark Lagrangian reads ${\cal L}_V = \sum_i \bar q_i \gamma^\mu\left(i\p_\mu + \frac{1}{N_c}A_{B\mu}\right)q_i$. We will set $A_{B\mu} = (\mu_B,0,0,0)$ to introduce the chemical potential term
${\cal L}_{\rm chem} = \mu_B n_B$ with $n_B = \frac{1}{N_c}\sum_i\bar q_i\gamma^0q_i$. Equivalently, the Hamiltonian contains ${\cal H}_{\rm chem} = - \mu_B n_B$, which has the correct sign for the baryon chemical potential.
The $\text{U}(1)_B$ gauge transformation is $q \to e^{iQ^B\tilde\beta}q$ and $A_{B\mu} \to A_{B\mu} + \p_\mu \tilde\beta$.
Similarly, the $\text{U}(1)_{\text{EM}}$ gauge field $A_{Q\mu}$ can be introduced by
\begin{eqnarray}
    A_\mu = eQ A_{Q\mu}\,,
\end{eqnarray}
where the charge matrices are given by
%$
%Q^B = \frac{1}{3}{\bf 1}_{N_f}
%$,
$
Q = {\rm diag}\left(\frac{2}{3},-\frac{1}{3}\right)
= \frac{1}{6} + \frac{\tau_3}{2}
$ 
for $N_f=2$ and
$
Q = {\rm diag}\left(\frac{2}{3},-\frac{1}{3},-\frac{1}{3}\right)
= \frac{\lambda_3}{2} + \frac{\lambda_8}{2\sqrt3}
$ for $N_f=3$.

The anomalous $\text{U}(1)_A$ symmetries and the corresponding currents are given by
\begin{eqnarray}
    q_i \to e^{i\frac{\alpha_i}{2}Q_i^5\gamma_5}q_i\,,\quad
    j_\mu^A = \sum_{i=1}^{N_f}\frac{Q^5_i}{2}\bar q_i\gamma_5\gamma^\mu q_i\,.
\end{eqnarray}
Here, $Q_i^5$ is the $\text{U}(1)_A$ charge: for example, $(Q_u^5,Q_d^5,Q_s^5) = (1,1,1)$ for $\text{U}(1)_{\eta'}$, $(1,-1,0)$ for $\text{U}(1)_{\pi^0}$, and $\left(\frac{1}{\sqrt3},\frac{1}{\sqrt3},-\frac{2}{\sqrt3}\right)$
for $\text{U}(1)_\eta$ in the case of $N_f=3$,
while $(Q_u^5,Q_d^5) = (1,1)$ for $\text{U}(1)_{\eta'}$ and
$(1,-1,0)$ for $\text{U}(1)_{\pi^0}$ in the case of $N_f=2$.
The transformation law of the $\text{U}(1)_A$ NG boson is defined by
\begin{equation}
    \begin{aligned}
    &\Sigma = \exp\left(i\sum_{A=0,3,8}Q^5_A\varphi_A\right)
    = e^{i\varphi_0}\exp\left(i\sum_{A=3,8}Q^5_A\varphi_A\right)\,,\\
    &\varphi_A \to \varphi_A + q_A\alpha\,,
    \end{aligned}
\end{equation}
where $q_A$ is the $\text{U}(1)_A$ charge of the condensate. If the condensate is $q\bar q$ (hadronic matter), then $q_A = 1$. If it is $qq\bar q\bar q$, then $q_A=2$ (dense quark matter).
The conservation law of the $\text{U}(1)_A$ current is violated by triangle anomalies
\begin{eqnarray}
    \partial^\mu j_\mu^A = - \frac{N_c}{16\pi^2}\sum_{i=1}^{N_f} Q_i^5 V_{i\,\mu\nu} \tilde V^{\mu\nu}_i\,,
\end{eqnarray}
with $V_{i\,\mu\nu} = \p_\mu A_{i\,\nu} - \p_\nu A_{i\,\mu}$ and
$\tilde V^{\mu\nu} = \frac{1}{2}\epsilon^{\mu\nu\alpha\beta}V_{\alpha\beta}$.

The effective theory should reproduce the same anomaly relation.
$\p^\mu j_\mu^A$ can be found in the effective theory from the change of the effective action under the $\text{U}(1)_A$ transformation as $\delta^A S = - \int d^4x\, (\p^\mu\alpha) j_\mu^A = \int d^4x\, \alpha \p^\mu j_\mu^A$\,\cite{Son:2004tq}. Therefore, the anomaly part of the effective Lagrangian should take the following form:
\begin{eqnarray}
    {\cal L}_{\rm anom} = \frac{N_c}{8\pi^2 q_A} \p_\mu\varphi_A \sum_{i=1}^{N_f}Q_i^5 V_{i\,\nu}\tilde V_i^{\mu\nu}\,.
    \label{eq:L_anom_0}
\end{eqnarray}
It is straightforward to verify that
\begin{eqnarray}
    \delta {\cal L}_{\rm anom} %= \frac{N_c}{8\pi^2}\p_\mu\alpha\sum_{i=1}^{N_f}Q_i^5V_{i\,\nu}\tilde V_i^{\mu\nu}
    = \alpha\left(- \frac{N_c}{16\pi^2} \sum_{i=1}^{N_f}Q_i^5 V_{i\,\mu\nu}\tilde V_i^{\mu\nu}\right)\,,
\end{eqnarray}
up to a total derivative term.

The gauge fields $A_{i\,\mu}$ include both the physical EM gauge field and the spurion gauge fields ($A_{B\mu}$ and $A_{I\mu}$), which are non-dynamical. We will keep $A_{i\,\mu}$ as general as possible during intermediate calculations, and at the very end we will substitute a concrete background configuration.

\subsubsection{$A,B,\gamma$}

Let us consider a concrete example with the EM gauge field $A_{Q\mu}$ and the baryon spurion gauge field $A_{B\mu}$ as
\begin{eqnarray}
    A_\mu = eQA_{Q\mu} - Q^BA_{B\mu}\,.
\end{eqnarray}
Plugging this into Eq.~(\ref{eq:L_anom_0}), we find
\begin{equation}
    \begin{aligned}
    {\cal  L}_{\rm anom} &= \frac{1}{8\pi^2q_A}\p_\mu\varphi_A
    \bigg[
    e^2C_{A\gamma\gamma} A_{Q\nu}\tilde F^{\mu\nu}\\
    &- e C_{AB\gamma}\left(
    A_{Q\nu}\tilde B^{\mu\nu} + A_{B\nu} \tilde F^{\mu\nu}
    \right)
    + C_{ABB} A_{B\nu} \tilde B^{\mu\nu}
    \bigg]\,
    \label{eq:anom_ABG_sym}
    \end{aligned}
\end{equation}
where we have $F_{\mu\nu} = \p_\mu A_{Q\nu} - \p_\nu A_{Q\mu}$ and
$B_{\mu\nu} = \p_\mu A_{B\nu} - \p_\nu A_{B\mu}$. We perform integration by parts to replace $A_{Q\nu}\tilde B^{\mu\nu} + A_{B\nu}\tilde F^{\mu\nu} \to 2 A_{B\nu}\tilde F^{\mu\nu}$. This yields
\begin{equation}
    \begin{aligned}
    {\cal  L}_{\rm anom} &= \frac{1}{8\pi^2q_A}\p_\mu\varphi_A
    \bigg[
    e^2C_{A\gamma\gamma} A_{Q\nu}\tilde F^{\mu\nu}
    - 2e C_{AB\gamma}A_{B\nu} \tilde F^{\mu\nu} \\
    &+ C_{ABB} A_{B\nu} \tilde B^{\mu\nu}
    \bigg]\,.
    \label{eq:anom_ABG_asym}
    \end{aligned}
\end{equation}
The anomaly coefficients are given by
\begin{equation}
    \begin{aligned}
    &C_{A\gamma\gamma} = N_c \sum_{i=1}^{N_f}Q_i^5 Q_i^2\,,\\
    &C_{AB\gamma} = N_c \sum_{i=1}^{N_f}Q_i^5Q^B_i Q_i\,,\\
    &C_{ABB} = N_c \sum_{i=1}^{N_f}Q_i^5(Q^B_i)^2\,.
    \end{aligned}
\end{equation}
These agree with the results in Ref.~\cite{Son:2004tq}.

\subsubsection{$A,B, I, \gamma$}

We here extend the expression for ${\cal L}_{\rm anom}$ given in Ref.~\cite{Son:2004tq} by introducing the background spurion field $A_{I\mu}$ for the $\text{U}(1)_I$ isospin as
\begin{eqnarray}
    A_\mu = eQ A_{Q\mu} - Q^BA_{B\mu} - Q^I A_{I\mu}\,,
\end{eqnarray}
with $Q^I_u = \frac{1}{2}$, $Q^I_d = -\frac{1}{2}$, and $Q_s^I = 0$.
Plugging this into Eq.~(\ref{eq:L_anom_0}) as before, we find
the anomaly Lagrangian including $A_{I\mu}$ as
\begin{equation}
    \begin{aligned}
    {\cal L}_{\rm anom} &= \frac{1}{8\pi^2 q_A}\partial_\mu \varphi_A
    \bigg[
    e^2 C_{A\gamma\gamma} A_{Q\nu} \tilde F^{\mu\nu}\\
    &- e C_{AB\gamma} \left(A_{B\nu} \tilde F^{\mu\nu} + A_{Q\nu} \tilde B^{\mu\nu}\right)\\
    &+ C_{ABB} A_{B\nu} \tilde B^{\mu\nu}
    + C_{ABI} \left(A_{B\nu} \tilde I^{\mu\nu}
    + A_{I\nu} \tilde B^{\mu\nu}\right)\\
    &+ 
    C_{AII} A_{I\nu} \tilde I^{\mu\nu}
    -e C_{AI\gamma} \left(A_{I\nu} \tilde F^{\mu\nu} + A_{Q\nu} \tilde I^{\mu\nu}\right)
    \bigg]\,.
    \label{eq:L_anom}
    \end{aligned}
\end{equation}
We again perform integration by parts to obtain the following:
\begin{equation}
    \begin{aligned}
    {\cal L}_{\rm anom} &= \frac{1}{8\pi^2 q_A}\partial_\mu \varphi_A
    \bigg[
    e^2 C_{A\gamma\gamma} A_{Q\nu} \tilde F^{\mu\nu}
    - 2e C_{AB\gamma} A_{B\nu} \tilde F^{\mu\nu}
    \\
    & + C_{ABB} A_{B\nu} \tilde B^{\mu\nu}
    + C_{ABI} \left(A_{B\nu} \tilde I^{\mu\nu}
    + A_{I\nu} \tilde B^{\mu\nu}\right)
    \\
    & + 
    C_{AII} A_{I\nu} \tilde I^{\mu\nu}
    -2 e C_{AI\gamma} A_{I\nu} \tilde F^{\mu\nu} 
    \bigg]\,,
    \label{eq:L_anom_asym}
    \end{aligned}
\end{equation}
with $I_{\mu\nu} = \p_\mu A_{I\nu} - \p_\nu A_{I\mu}$ and the additional anomaly coefficients:
\begin{equation}
    \begin{aligned}
    %C_{A\gamma\gamma} = N_c \sum_{i=1}^{N_f}Q_i^5 Q_i^2\,,\quad
    %C_{AB\gamma} = N_c\sum_{i=1}^{N_f}Q_i^5Q^B_i Q_i\,,\quad
    &C_{AI\gamma} = N_c\sum_{i=1}^{N_f}Q_i^5 Q_i^I Q_i\,,\\
    %C_{ABB} = N_c \sum_{i=1}^{N_f}Q_i^5(Q^B_i)^2\,,\quad
    &C_{AII} = N_c\sum_{i=1}^{N_f}Q_i^5 (Q_i^I)^2\,,\\
    &C_{ABI} = N_c\sum_{i=1}^{N_f}Q_i^5Q^B_i Q_i^I\,.
    \label{eq:Cs_I}
    \end{aligned}
\end{equation}

Note that, although we have introduced $A_{I\mu}$ in addition to $A_{Q\mu}$ and $A_{B\mu}$,
the number of independent degrees of freedom is two for the case $N_f=2$, since
$A_{u\,\mu} = \frac{2e}{3}A_{Q\mu} - \frac{1}{3}A_{B\mu} - \frac{1}{2}A_{I\mu}$
and
$A_{d\,\mu} = -\frac{e}{3}A_{Q\mu} - \frac{1}{3}A_{B\mu} + \frac{1}{2}A_{I\mu}$ (we have put $N_c=3$).
The charge matrix for $N_f=2$ obeys the Gell-Mann–Nishijima formula $Q = \frac{1}{2}Q^B + Q^I$. Substituting $Q^I = Q - \frac{1}{2}Q^B$ into Eq.~(\ref{eq:Cs_I}), one finds that the additional anomaly coefficients ($C_{AI\gamma},C_{ABI},C_{AII}$) can be expressed as the following linear combinations:
\begin{eqnarray}
    C_{AI\gamma} = C_{A\gamma\gamma} - \frac{1}{2}C_{AB\gamma}\,,\quad
    C_{ABI} = C_{AB\gamma} - \frac{1}{2}C_{ABB}\,,\label{eq:C_Nf2_1}\\
    C_{AII} = C_{A\gamma\gamma} - C_{AB\gamma} + \frac{1}{4}C_{ABB}\,.\qquad\qquad
    \label{eq:C_Nf2_2}
\end{eqnarray}
Equivalently, one can achieve the same result by shifting the fields as $A_{Q\mu} \to A_{Q\mu} - \frac{A_{I\mu}}{e}$ and $A_{B\mu} \to A_{B\mu} - \frac{A_{I\mu}}{2}$ in the expression for ${\cal L}_{\rm anom}$ in Eq.~(\ref{eq:anom_ABG_sym}) where $A_{I\mu}$ is absent.

\subsubsection{Anomalous currents}

One of the important effects related to the WZW-like Lagrangian shown above arises when the following three conditions simultaneously hold: 1) a spatial modulation of the NG bosons such that $\p_\mu \varphi_A \neq 0$, 2) the presence of external background fields, and 3) nonzero chemical potentials.
Here we give a generic formula for the anomalous currents derived from Eq.~(\ref{eq:L_anom_asym}) via
$j_X^\mu \equiv \frac{\delta {\cal L}_{\rm anom}}{\delta X_\mu} - \p_\nu\frac{\delta{\cal L}_{\rm anom}}{\delta \p_\nu X_\mu}$:
\begin{equation}
    \begin{aligned}
    j_{\gamma}^\mu 
    &= \frac{1}{4\pi^2q_A} \p_\nu\varphi_A
    \left(
    e^2 C_{A\gamma\gamma} \tilde F^{\nu\mu}
    - e C_{AB\gamma} \tilde B^{\nu\mu}
    - e C_{AI\gamma} \tilde I^{\nu\mu}
    \right)\,,%\label{eq:J_g1}
    \\
    j_{B}^\mu 
    &= \frac{1}{4\pi^2q_A} \p_\nu\varphi_A
    \left(
    - e C_{AB\gamma} \tilde F^{\nu\mu}
    + C_{ABB} \tilde B^{\nu\mu}
    + C_{ABI} \tilde I^{\nu\mu}
    \right)\,,
    \label{eq:J_B1}\\
    j_{I}^\mu 
    &= \frac{1}{4\pi^2q_A} \p_\nu\varphi_A
    \left(
    - e C_{AI\gamma} \tilde F^{\nu\mu}
    + C_{ABI} \tilde B^{\nu\mu}
    + C_{AII} \tilde I^{\nu\mu}
    \right)
    %\label{eq:J_I1}
    \,.
    \end{aligned}
\end{equation}

For the $N_f=2$ case, due to the relations (\ref{eq:C_Nf2_1}) and (\ref{eq:C_Nf2_2}), these currents obey the following relation, which is the current version of the Gell-Mann–Nishijima formula:
\begin{eqnarray}
    j_\gamma^\mu = - e\left(\frac{1}{2}j_{B}^\mu + j_I^\mu \right)\,.
\end{eqnarray}

\subsection{Examples: Spatially modulated mesons}

Here, we will show various examples of spatially modulated mesons under a magnetic field and rotation.
We set the background fields as $\tilde F^{i0} = - B_i$, $A_B^\mu = \mu_B n^\mu$,
and $A_I^\mu = \mu_I n^\mu$ with $n^\mu = (1,\vec v)$, where $\vec\nabla \times \vec v = 2\vec\Omega$. This leads to $\tilde B^{i0} = 2\mu_B \Omega_i$ and $\tilde I^{i0} = 2\mu_I \Omega_i$. We will use the notations $\vec B = (B_1,B_2,B_3)$ and $\vec\Omega = (\Omega_1,\Omega_2,\Omega_3)$ in the following.

\subsubsection{$N_f = 2$ at low density}

There are two $\text{U}(1)_A$ symmetries: one is $\text{U}(1)_0$ and the other is $\text{U}(1)_3$.
The NG boson for the former is $\varphi_0$, corresponding to the $\eta'$ meson,
whereas the latter is $\varphi_3$, corresponding to the neutral pion $\pi^0$. We take $Q^5_{\pi^0} = \tau^3$ and $Q^5_{\eta'} = {\bf 1}_2$, and
the anomaly coefficients are summarized in Table~\ref{tab:NF=2_low}.
\begin{table}[ht]
    \centering
    \begin{tabular}{c|cccccc}
    & $C_{A\gamma\gamma}$ & $C_{AB\gamma}$ & $C_{ABB}$ & $C_{AI\gamma}$ & $C_{AII}$ & $C_{ABI}$\\
    \hline
    $\eta'$  & $\frac{5}{3}$ & $\frac{1}{3}$ & $\frac{2}{3}$ & $\frac{3}{2}$ & $\frac{3}{2}$ & $0$\\
    $\pi^0$ & $1$ & $1$ & $0$ & $\frac{1}{2}$ & $0$ & $1$
    \end{tabular}
    \caption{$N_f = 2$ at low density}
    \label{tab:NF=2_low}
\end{table}
Using these coefficients for ${\cal L}_{\rm anom}$ in Eq.~(\ref{eq:L_anom_asym}), we find the WZW-like anomaly Lagrangians:
\begin{equation}
    \begin{aligned}
    {\cal L}_{\rm anom}^{(\eta')} 
    &= \frac{e}{4\pi^2}\left(\frac{\mu_B}{3}+\frac{3\mu_I}{2}\right)\vec B \cdot \vec\nabla\varphi_0
    + \frac{1}{4\pi^2}\left(\frac{2\mu_B^2}{3}+\frac{3\mu_I^2}{2}\right) \vec \Omega \cdot \vec\nabla\varphi_0\label{eq:L_anom_eta_Nf=2} \\
    &= \frac{e}{4\pi^2f_{\eta'}}\left(\frac{\mu_B}{3}+\frac{3\mu_I}{2}\right)\vec B \cdot \vec\nabla\eta'
    + \frac{1}{4\pi^2f_{\eta'}}\left(\frac{2\mu_B^2}{3}+\frac{3\mu_I^2}{2}\right) \vec \Omega \cdot \vec\nabla\eta'\,,
    \end{aligned}
\end{equation}
and
\begin{equation}
    \begin{aligned}
    {\cal L}_{\rm anom}^{(\pi^0)} 
    &= \frac{e}{4\pi^2}\left(\mu_B+\frac{\mu_I}{2}\right)\vec B \cdot \vec\nabla\varphi_3
    + \frac{1}{2\pi^2}\mu_B \mu_I \vec \Omega \cdot \vec\nabla\varphi_3\\
    &= \frac{e}{4\pi^2f_\pi}\left(\mu_B+\frac{\mu_I}{2}\right)\vec B \cdot \vec\nabla\pi^0
    + \frac{1}{2\pi^2f_\pi}\mu_B \mu_I \vec \Omega \cdot \vec\nabla\pi^0\,,
    \label{eq:L_anom_pi0_Nf2}
    \end{aligned}
\end{equation}
where we introduced the physical fields 
$\varphi_3 = \frac{\pi^0}{f_\pi}$ and $\varphi_0 = \frac{\eta'}{f_{\eta'}}$, which are canonically normalized as
${\cal L}_{\rm ChPT} = \frac{f_\pi^2}{4} {\rm Tr}D_\mu\hat\Sigma^\dag D^\mu\hat\Sigma + \frac{f_{\eta'}^2}{4}{\rm Tr}\p_\mu e^{-i\varphi_0{\bf 1}_2}\p^\mu e^{i\varphi_0{\bf 1}_2} + \cdots = \frac{1}{2}(\p_\mu\pi^0)^2 + \frac{1}{2}(\p_\mu\eta')^2 + \cdots$ with the definition $\hat \Sigma = e^{i\varphi_3\tau_3}$.
The terms proportional to $\vec B$ are partially studied in Refs.~\cite{Son:2004tq,Son:2007ny,Qiu:2023guy,Gronli:2022cri}, while
those proportional to $\vec\Omega$ are partially studied in 
Refs.~\cite{Huang:2017pqe,Eto:2021gyy,Eto:2023rzd,Eto:2023tuu}.

\subsubsection{$N_f = 3$ at low density}

There are three $\text{U}(1)_A$ symmetries for $N_f=3$:
$\text{U}(1)_0$, $\text{U}(1)_3$, and $\text{U}(1)_8$ for $\eta'$, $\pi^0$, and $\eta$, respectively.
We take $Q^5_{\pi^0} = \lambda^3$, $Q^5_{\eta} = \lambda_8$, and $Q^5_{\eta'} = {\bf 1}_3$, and the anomaly coefficients are summarized in Table~\ref{tab:NF=3_low}.
\begin{table}[ht]
    \centering
    \begin{tabular}{c|cccccc}
    & $C_{A\gamma\gamma}$ & $C_{AB\gamma}$ & $C_{ABB}$ & $C_{AI\gamma}$ & $C_{AII}$ & $C_{ABI}$\\
    \hline
    $\eta'$ & $2$ & $0$ & $1$ & $\frac{3}{2}$ & $\frac{3}{2}$ & $0$\\
    $\eta$ & $\frac{1}{\sqrt3}$ & $\frac{1}{\sqrt3}$ & $0$ & $\frac{\sqrt3}{2}$ & $\frac{\sqrt3}{2}$ & $0$\\
    $\pi^0$ & $1$ & $1$ & $0$ & $\frac{1}{2}$ & $0$ & $1$
    \end{tabular}
    \caption{$N_f = 3$ at low density}
    \label{tab:NF=3_low}
\end{table}
Then the WZW-like anomaly Lagrangians are
\begin{equation}
    \begin{aligned}
    {\cal L}_{\rm anom}^{(\pi^0)} 
    &= \frac{e}{4\pi^2}\left(\mu_B+\frac{\mu_I}{2}\right)\vec B \cdot \vec\nabla\varphi_3
    + \frac{\mu_B \mu_I}{2\pi^2} \vec \Omega \cdot \vec\nabla\varphi_3\\
    &= \frac{e}{4\pi^2f_\pi}\left(\mu_B+\frac{\mu_I}{2}\right)\vec B \cdot \vec\nabla\pi^0
    + \frac{\mu_B \mu_I}{2\pi^2f_\pi} \vec \Omega \cdot \vec\nabla\pi^0\,.
    \label{eq:Lanom_Nf3_pi0}
    \end{aligned}
\end{equation}
\begin{equation}
    \begin{aligned}
    {\cal L}_{\rm anom}^{(\eta)} 
    &= \frac{e}{4\pi^2}\left(\frac{\mu_B}{\sqrt3}+\frac{\sqrt3\mu_I}{2}\right)\vec B \cdot \vec\nabla\varphi_8
    + \frac{\sqrt3\mu_I^2}{8\pi^2} \vec \Omega \cdot \vec\nabla\varphi_8 \\
    &= \frac{e}{4\pi^2f_\pi}\left(\frac{\mu_B}{\sqrt3}+\frac{\sqrt3\mu_I}{2}\right)\vec B \cdot \vec\nabla\eta
    + \frac{\sqrt3\mu_I^2}{8\pi^2f_\pi} \vec \Omega \cdot \vec\nabla\eta\,,
    \label{eq:Lanom_Nf3_eta}
    \end{aligned}
\end{equation}
\begin{equation}
    \begin{aligned}
    {\cal L}_{\rm anom}^{(\eta')} 
    &= \frac{3e \mu_I}{8\pi^2} \vec B \cdot \vec\nabla\varphi_0
    + \frac{1}{4\pi^2}\left(\mu_B^2+\frac{3\mu_I^2}{2}\right) \vec \Omega \cdot \vec\nabla\varphi_0\\
    &= \frac{\sqrt6 e \mu_I}{8\pi^2f_{\eta'}} \vec B \cdot \vec\nabla\eta'
    + \frac{\sqrt6}{12\pi^2f_{\eta'}}\left(\mu_B^2+\frac{3\mu_I^2}{2}\right) \vec \Omega \cdot \vec\nabla\eta'\,,
    \label{eq:Lanom_Nf3_eta'}
    \end{aligned}
\end{equation}
where we introduced the physical fields $\varphi_0 = \sqrt{\frac{2}{3}}\frac{\eta'}{f_{\eta'}}$, $\varphi_8 = \frac{\eta}{f_\pi}$, and $\varphi_3 = \frac{\pi^0}{f_{\pi}}$, which are canonically normalized as
${\cal L}_{\rm ChPT} = \frac{f_\pi^2}{4} {\rm Tr}D_\mu\hat\Sigma^\dag D^\mu\hat\Sigma + \frac{f_{\eta'}^2}{4}{\rm Tr}\p_\mu e^{-i\varphi_0{\bf 1}_3}\p^\mu e^{i\varphi_0{\bf 1}_2} + \cdots = \frac{1}{2}(\p_\mu\pi^0)^2 + \frac{1}{2}(\p_\mu\eta)^2 + \frac{1}{2}(\p_\mu\eta')^2 + \cdots$ with $\hat \Sigma = e^{i\varphi_3\lambda_3+i\varphi_8\lambda_8}$.
The term proportional to $\vec B$ in ${\cal L}_{\rm anom}^{(\pi^0)}$ is studied in Ref.~\cite{Son:2007ny}, and
the terms proportional to $\vec \Omega$ are partially studied in Ref.~\cite{Nishimura:2020odq}.

\subsubsection{$N_f = 2$ at high density: 2SC}

There is only the $\eta'$ meson in the 2SC phase\footnote{It is also called the $\eta$ boson in \cite{Son:2004tq, Son:2007ny}, since it is not heavy owing to the suppression of instanton effects at high density.
} because the condensate $X^a \sim \epsilon^{abc}\epsilon_{ij}q_{Li}^bq_{Lj}^c$ is invariant under $\text{SU}(2)_f$. The $Q_5$ charge is $(Q^5_u,Q^5_d)=(1,1)$ and the anomaly coefficients are summarized in Table~\ref{tab:NF=2_high}.
\begin{table}[ht]
    \centering
    \begin{tabular}{c|cccccc}
    & $C_{A\gamma\gamma}$ & $C_{AB\gamma}$ & $C_{ABB}$ & $C_{AI\gamma}$ & $C_{AII}$ & $C_{ABI}$\\
    \hline
    $\eta'$  & $\frac{5}{3}$ & $\frac{1}{3}$ & $\frac{2}{3}$ & $\frac{3}{2}$ & $\frac{3}{2}$ & $0$
    \end{tabular}
    \caption{$N_f = 2$ at high density}
    \label{tab:NF=2_high}
\end{table}
The WZW-like anomaly Lagrangian is given by
\begin{equation}
    {\cal L}_{\rm anom}^{(\eta')} 
    = \frac{e}{8\pi^2}\left(\frac{\mu_B}{3}+\frac{3\mu_I}{2}\right)\vec B \cdot \vec\nabla\varphi_0
    + \frac{1}{8\pi^2}\left(\frac{2\mu_B^2}{3}+\frac{3\mu_I^2}{2}\right) \vec \Omega \cdot \vec\nabla\varphi_0\,.
    \label{eq:L_anom_2SC}
\end{equation}
Note that this is  half of the low-density expression for ${\cal L}_{\rm anom}^{(\eta')}$ due to $q_A = 2$ in Eq.~(\ref{eq:L_anom_eta_Nf=2}), reflecting $q_A = 2$.
The first term with $\mu_I = 0$ is studied in Ref.~\cite{Son:2004tq}.\footnote{Eq.~(43) of \cite{Son:2007ny} also includes the same term, but with the different coefficient $\frac{1}{36\pi^2}$.}

\subsubsection{$N_f = 3$ at high density: CFL}

When $m_s > m_{u,d}$, the lightest CFL meson is $\varphi_{\bar ss}$, which lives in the third diagonal element of $\Sigma_{ij} = X_i^\dag Y_j \sim \epsilon_{ikl}\epsilon_{jmn}(q_{Lk}q_{Ll})^*(q_{Rm}q_{Rn})$. $\text{U}(1)_{\bar ss}$ shifts only $\varphi_{\bar ss}$, so the corresponding charge is $(Q_5^u,Q_5^d,Q_5^s) = (1,1,-1)$. On the other hand, if we take the asymptotically high-density limit $m_u = m_d = m_s$, the mixing between mesons vanishes. Then, the $\eta'$ meson is associated with the charge $(Q_5^u,Q_5^d,Q_5^s) = (1,1,1)$.
The anomaly coefficients are summarized in Table~\ref{tab:NF=3_high}.
\begin{table}[ht]
    \centering
    \begin{tabular}{c|cccccc}
    &  $C_{A\gamma\gamma}$ & $C_{AB\gamma}$ & $C_{ABB}$ & $C_{AI\gamma}$ & $C_{AII}$ & $C_{ABI}$\\
    \hline
    $\varphi_{\bar ss} $ & $\frac{4}{3}$ & $\frac{2}{3}$ & $\frac{1}{3}$ & $\frac{3}{2}$ & $\frac{3}{2}$ & $0$\\
    $\eta'$  & $2$ & $0$ & $1$ & $\frac{3}{2}$ & $\frac{3}{2}$ & $0$
    \end{tabular}
    \caption{$N_f = 3$ at high density}
    \label{tab:NF=3_high}
\end{table}

Thus the WZW-type anomaly Lagrangian of the $\bar ss$ meson for $m_s > m_{u,d}$ reads
\begin{equation}
    {\cal L}_{\rm anom}^{(\bar ss)} = \frac{e}{8\pi^2}\left(\frac{2\mu_B}{3}+\frac{3\mu_I}{2}\right)\vec B \cdot \vec\nabla \varphi_{\bar ss} + \frac{1}{8\pi^2}\left(\frac{\mu_B^2}{3} + \frac{3\mu_I^2}{2}\right) \vec \Omega \cdot \vec\nabla \varphi_{\bar ss}\,.
    \label{eq:Lag_anom_ss}
\end{equation}
The first term with $\mu_I=0$ is given in Ref.~\cite{Son:2007ny}.

Similarly, the anomaly term of the $\eta'$ meson for the case $m_u=m_d=m_s$ at the asymptotically high density limit reads
\begin{eqnarray}
    {\cal L}_{\rm anom}^{(\eta')} = \frac{3e\mu_I}{16\pi^2}\vec B \cdot \vec\nabla \varphi_0 + \frac{1}{8\pi^2}\left(\mu_B^2 + \frac{3\mu_I^2}{2}\right)\vec\Omega \cdot \vec\nabla \varphi_0\,.
    \label{eq:L_anom_eta'_CFL}
\end{eqnarray}
The second term with $\mu_I=0$ coincides with the result in Ref.~\cite{Nishimura:2020odq} with $\varphi_0 = \sqrt{\frac{2}{3}}\frac{\eta'}{f_{\tilde \eta'}}$.

\subsection{Anomalous currents on the topological solitons}

Finally, we briefly discuss the anomalous effects in the presence of topological solitons~\cite{Son:2004tq,Fukushima:2018ohd}.

\subsubsection{Anomalous charges on an axial domain wall}
\label{sec:anomalous_charges}
Let us consider the simplest case where there exists a
homogeneous and constant magnetic background field $F_{12} = - \tilde F^{30} = B_z$, and rigid rotation around the $z$ axis. The rotation is incorporated via $A_{B\mu} = \mu_B n_\mu$ and $A_{I\mu} = \mu_I n_\mu$ with $n^\mu = (1,\vec v)$, where $\vec\nabla \times \vec v = 2 \vec \Omega$ and 
$\vec \Omega = (0,0,\Omega)$, leading to
$\tilde B^{30} = 2\mu_B\Omega$ and $\tilde I^{30} = 2 \mu_I \Omega$.
Then, domain walls perpendicular to the $z$ axis spontaneously arise to form the so-called chiral soliton lattice. Namely, we assume that $\p_\mu \varphi_A \propto \delta^3_\mu$.
Substituting these into Eq.~(\ref{eq:J_B1}), we find the following anomalous charge densities induced around the domain walls:
\begin{align}
    j_{\gamma}^0 
    &= \frac{1}{4\pi^2q_A} \p_3\varphi_A
    \left(
    - e^2 C_{A\gamma\gamma} B_z
    - 2e C_{AB\gamma} \mu_B \Omega
    - 2 C_{AI\gamma} \mu_I \Omega
    \right)\,,\\
    j_{B}^0 
    &= \frac{1}{4\pi^2q_A} \p_3\varphi_A
    \left(
    e C_{AB\gamma} B_z
    + 2 C_{ABB} \mu_B \Omega
    + 2 C_{ABI} \mu_I \Omega
    \right)\,,\\
    j_{I}^0 
    &= \frac{1}{4\pi^2q_A} \p_3\varphi_A
    \left(
    e C_{AI\gamma} B_z
    + 2 C_{ABI} \mu_B \Omega
    + 2 C_{AII} \mu_I \Omega
    \right)\,.
\end{align}
Assuming $\varphi_A$ changes by $2\pi$ when crossing the domain wall, the baryon charge per unit area of the $\{\pi^0,\eta,\eta'\}$-domain wall reads 
\begin{eqnarray}
    N_B^{(A)} = \frac{
    e C_{AB\gamma} B_z
    + 2 C_{ABB} \mu_B \Omega
    + 2 C_{ABI} \mu_I \Omega
    }{2\pi q_A}\,.
\end{eqnarray}
This agrees with the $\pi^0$ CSL under a magnetic field \cite{Son:2004tq,Son:2007ny}.
Furthermore, the above charge formula implies that a magnetic moment orthogonal to the domain wall is induced on the $\{\pi^0,\eta,\eta'\}$-domain wall:
\begin{eqnarray}
    M_z^{(A)} = \frac{e C_{AB\gamma}\mu_B + e C_{AI\gamma}\mu_I}{2\pi q_A}\,.
\end{eqnarray}

\subsubsection{Anomalous electric currents on an axial vortex}

Another interesting effect appears when there is a quantum vortex of the NG fields \cite{Son:2004tq}. When the axial NG boson $\varphi_A$ winds nontrivially around a cycle in an inner space, namely when there is a vortex, we should keep $[\p_\mu,\p_\nu]\varphi_A \neq 0$ in general.
Then, there is an additional contribution to the electric current in addition to those given in Eqs.~(\ref{eq:J_B1}). Calculating carefully 
$j_X^\mu \equiv \frac{\delta {\cal L}_{\rm anom}}{\delta X_\mu} - \p_\nu\frac{\delta{\cal L}_{\rm anom}}{\delta \p_\nu X_\mu}$ with
${\cal L}_{\rm anom}$ given in Eq.~(\ref{eq:L_anom_asym}), we find the following terms:
\begin{equation}
    j_{\gamma,\text{vor}}^\mu 
    %&=& 
    =
    \frac{1}{4\pi^2q_A} \epsilon^{\mu\nu\alpha\beta}
    \left(
    e^2 C_{A\gamma\gamma} A_\nu
    - e C_{AB\gamma} B_\nu
    - e C_{AI\gamma} I_\nu
    \right)\p_\alpha\p_\beta\varphi_A\,.
    \label{eq:J_g2}
\end{equation}
To be concrete, let us assume a straight line vortex satisfying $[\p_1,\p_2]\varphi_A = 2\pi \delta^{(2)}(x,y)$ along the $z$ axis. Furthermore, we assume nonzero baryon and isospin chemical potentials such that $B_\mu = \mu_B\delta^0_\mu$ and $I_\mu = \mu_I\delta^0_\mu$, while the EM gauge potential is turned off ($A_\mu = 0$). Then, the anomalous electric current density flowing along the vortex line is given by
\begin{eqnarray}
    j_{\gamma,\text{vor}}^{\mu=3} 
    = \frac{1}{2\pi q_A} 
    \left(
    e C_{AB\gamma} \mu_B
    + e C_{AI\gamma} \mu_I
    \right)\delta^{(2)}(x,y)\,,
\end{eqnarray}
and therefore the total electric current reads
\begin{eqnarray}
    J_{\gamma,\text{vor}}
    = \frac{e C_{AB\gamma} \mu_B
    + e C_{AI\gamma} \mu_I}{2\pi q_A}.
\end{eqnarray}

\section{Conclusion and discussions} \label{sec:conclusion}

In this work, we have investigated anomaly-induced WZW terms in dense QCD matter under external magnetic fields and rotation. By introducing electromagnetic, baryon-number, and isospin background gauge fields, we systematically derived the anomalous couplings of neutral mesons for both the $N_f=2$ and $N_f=3$ theories. The resulting effective Lagrangians contain characteristic topological terms proportional to $\vec{B}\cdot \vec{\nabla}\phi$ and $\vec{\Omega}\cdot \vec{\nabla}\phi$, where $\phi$ denotes neutral meson fields such as $\pi^0$, $\eta$, and $\eta'$.

Our analysis provides a unified description of anomaly-induced effects generated by magnetic fields and rotation. Although these two external backgrounds originate from different physical mechanisms, they enter the low-energy effective theory through closely related topological couplings. 
In particular, 
within the slow rotation approximation employed in this work,
rotation can be incorporated as a background gauge field associated with finite density, leading to anomaly-induced interactions analogous to those generated by magnetic fields.

We have also derived the corresponding anomalous electromagnetic, baryon-number, and isospin currents from the viewpoint of $\mathrm{U}(1)$ anomalies. Applying these results to domain walls and vortices of neutral mesons, we obtained the associated anomaly-induced charges and currents. These expressions provide a direct link between microscopic triangle anomalies and macroscopic topological excitations in dense matter.

The anomaly-induced terms obtained here provide the microscopic origin of various topological phases that have recently attracted considerable attention in dense QCD. These include chiral soliton lattices, $\eta'$ chiral soliton lattices, domain-wall Skyrmion phases, baryonic crystals, baryonic vortex phases, and related inhomogeneous states. Since the terms derived in this work are determined entirely by anomalies, they are universal and largely independent of microscopic details.

One caveat concerning the applicability of our results is that the correspondence between rotation and a background magnetic field employed here is restricted to the leading order in the slow-rotation expansion. Beyond this approximation, rotation introduces effects with no direct magnetic-field analogue, including an explicit dependence on the distance from the rotation axis. Lattice studies of rotating Yang--Mills theory and QCD have revealed nontrivial rotational modifications of their equilibrium properties \cite{Yamamoto:2013zwa,Braguta:2021jgn,Yang:2023vsw}, while spatially resolved lattice simulations have found radially inhomogeneous states and spatially separated confinement and deconfinement regions \cite{Chernodub:2022veq,Braguta:2023iyx}. Similar radial dependence has also been obtained in a semiclassical analysis beyond linear order in the angular velocity \cite{Jiang:2024zsw}. While the existence of the CSL discussed here is a leading-order effect in the angular velocity, higher-order rotational effects may modify its transverse homogeneity and induce a nontrivial radial structure. Investigating such corrections is beyond the scope of the present work and is left for future study.

Several future directions are worth pursuing. One important application is the systematic study of ground states in the simultaneous presence of magnetic fields, rotation, baryon density, and isospin density. It would also be interesting to investigate the holographic realization of the anomaly-induced terms derived here and their implications for strongly coupled QCD matter.

More broadly, our results support the emerging viewpoint that quantum anomalies do not merely induce currents on a given background but can actively determine the structure of the ground state itself. Understanding this anomaly-driven organization of matter may provide a unified framework for a wide variety of topological phases in dense QCD and related many-body systems.

One of the most promising directions for future work is investigating the CSL in the presence of a magnetic field and rotation simultaneously, and studying it in holographic QCD \cite{Amano:2025iwi}.

%%%%%%%%%%%%%%%%%%%%%%%%%%

\section*{Acknowledgements}
This work is supported in part by Japan Society for the Promotion of
 Science (JSPS) KAKENHI [Grants No. JP22H01221 and JP23K22492 (ME and
 MN), JP25K07324 (SS)] and the WPI program ``Sustainability with Knotted
 Chiral Meta Matter (WPI-SKCM$^2$)'' at Hiroshima University (ME and
 MN). 
 MA was provided funding by the National Institute of Technology, Oyama College.

\begin{appendix}

\section{Rotation as background U(1) gauge field} \label{sec:rotation}
In this section, we introduce the effects of rotation in terms of a background gauge field.
Just as how the chemical potential, $\mu$, is encoded in the temporal component of a background
$\mathrm{U}(1)$ gauge field, the angular velocity, $\Omega$, can also be encoded in the spatial components of a $\mathrm{U}(1)$ gauge field $A_i$ \cite{Son:2004tq, Abramchuk:2018jhd}.

First, let's start with a finite chemical potential background at rest, represented by a purely temporal $\mathrm{U}(1)$ gauge field, 
$A_{\mu} = (\mu,0,0,0)$.
We take the background to be rotating at constant velocity with respect to the laboratory frame.
We perform the Lorentz boost by the velocity $\vec{v} =  \vec{\Omega} \times \vec{x}$ where $\vec{\Omega}$ is a constant angular velocity in the laboratory frame\footnote{
A similar approach that introduces rotation via Lorentz boosts in a holographic QCD is found in Refs.~\cite{Zhao:2022uxc, Chen:2020ath}.
}.
The resulting gauge field is $A_\mu =\mu \gamma (v) (1, \vec{v})$
where $\gamma (v) = \frac{1}{\sqrt{1 - v^2}}$ is the Lorentz factor.
Assuming that $|v|$ is small, namely $|v| = |x| \Omega \ll c = 1$, 
then $\gamma \simeq 1 + \mathcal{O}(|v|^2)$. 
Then the gauge field can be given as
\begin{align}
A_{\mu} = \mu (1, \vec{\Omega} \times \vec{x}) + \mathcal{O}(|v|^2).
\label{eq:boosted_gauge}
\end{align}
In this frame, the magnetic field $\vec{H}$ associated with this gauge field \eqref{eq:boosted_gauge} is
\begin{align}
\vec{H} = \mu \vec{\nabla} \times (\vec{\Omega} \times \vec{x}) = 2 \mu \vec{\Omega}.
\label{eq:rotation_gauge}
\end{align}
Therefore, to leading order in the slow rotation expansion, 
the effect of rotation considered here can be represented by an effective background U(1) magnetic gauge field $\vec{H}= 2 \mu \vec{\Omega}$. 
This correspondence is useful for the anomaly-induced effects at linear order in $\Omega$ studied in this work, but should not be regarded as a general equivalence between rotation and a magnetic field.
This result is consistent with the formalism discussed in \cite{Son:2004tq} 
where rotated matter at finite density was described by the background magnetic gauge field $\vec{H} = 2 \mu \vec{\Omega}$.

There is another way to derive an equivalent identification.
Let's start with the Minkowski metric in cylindrical cooridnates,
\begin{align} 
ds^2 = 
dt^2 - dr^2 - r^2 d \Theta^2 - dz^2.
\end{align}
We then introduce rotation along the $z$-axis as a change of coordinates where $\Theta = \theta + \Omega t$. 
The metric becomes
\begin{align}
ds^2 = 
\left(
1 - r^2 \Omega^2
\right) dt^2
- dr^2 - r^2 d \theta^2 - 2 r^2 \Omega \, d t d \theta - d z^2.
\end{align}
or in Cartesian coordinates, the metric can be written as
\begin{align}
ds^2 = 
\Big(
1 - r^2 \Omega^2
\Big) dt^2
-
2 x \Omega \, dy dt
+
2 y \Omega \, dx dt
- dx^2 - dy^2 - dz^2.
\end{align}
This looks like the following four-dimensional Kaluza-Klein metric:
\begin{align}
g_{\mu \nu} &= 
\left(
\begin{array}{c|c}
1 - \mathcal{A}_k \mathcal{A}_k & - \mathcal{A}_j \\
\hline
- \mathcal{A}_i & - \delta_{ij}
\end{array}
\right),
\ \\
g^{\mu \nu} &= 
\left(
\begin{array}{c|c}
-1 & \mathcal{A}_j \\
\hline
\mathcal{A}_i & \delta_{ij} - \mathcal{A}_i \mathcal{A}_j
\end{array}
\right),
\notag \\
\det g &= -1,
\end{align}
where $\mathcal{A}_i = (-y \Omega, x \Omega, 0)$.
Let us now consider a massless complex scalar field, $\phi$, in this frame.
The Lagrangian is given by
\begin{align}
\mathcal{L}_{\mathrm{rot}} =& \
\sqrt{-g} g^{\mu \nu} \del_{\mu} \phi \del_{\nu} \bar{\phi}
\notag \\
=& \ 
\del_0 \phi \del_0 \bar{\phi}
-
\del_0 \phi \mathcal{A}_i \del_i \bar{\phi}
- \mathcal{A}_i \del_i \phi \del_0 \bar{\phi}
- 
\Big(
\delta_{ij} - \mathcal{A}_i \mathcal{A}_j
\Big) \del_i \phi \del_j \bar{\phi}.
\end{align}
In the case of the finite density, we perform the shift where $\del_0 \phi \to \del_0 \phi + i \mu \phi$.
%\begin{align}
%\del_0 \phi \to \del_0 \phi + i \mu \phi.
%\end{align}
%
The Lagrangian for static field where $\del_0 \phi = 0$ becomes
\if0
\begin{align}
\mathcal{L}_{\mathrm{rot}} 
\to 
\mathcal{L}_{\mathrm{d}}
=& \ 
\Big|
\del_0 \phi + i \mu \phi
\Big|^2
-
\Big(
\del_0 \phi + i \mu \phi
\Big)
\mathcal{A}_i \del_i \bar{\phi}
-
\mathcal{A}_i \del_i \phi
\Big(
\del_0 \bar{\phi} - i \mu \bar{\phi}
\Big)
-
\Big(
\delta_{ij} - \mathcal{A}_i \mathcal{A}_j
\Big) \del_i \phi \del_j \bar{\phi}.
\end{align}
If we consider $\del_0 \phi = 0$, then we have
\fi
\begin{align}
\mathcal{L}_{\mathrm{rot}} = 
\mu^2 |\phi|^2
-  i \mu \mathcal{A}_i \phi \del_i \bar{\phi}
+  i \mu \mathcal{A}_i \bar{\phi} \del_i \phi
- \delta_{ij} \del_i \phi \del_j \bar{\phi}
+ \mathcal{O} (\mathcal{A}^2).
\end{align}

On the other hand, if we had gauged the $\mathrm{U}(1)$ global symmetry where $\phi \to e^{i \theta} \phi$ in the non-rotating frame, the Lagrangian would have been given by
\begin{align}
\mathcal{L}_{\mathrm{g}}
= A_0^2 |\phi|^2
- i A_i \phi \del_i \bar{\phi}
+ i A_i \bar{\phi} \del_i \phi
-
\delta_{ij} 
\del_i \phi \del_j \bar{\phi},
\end{align}
where $A_i$ is the associated $\mathrm{U}(1)$ gauge field.
Comparing $\mathcal{L}_{\mathrm{rot}}$ and 
$\mathcal{L}_{\mathrm{g}}$, 
we find the following correspondence at $\mathcal{O} (\Omega)$ \cite{Son:2004tq, Abramchuk:2018jhd}:
\begin{align}
A_0 \ &\longleftrightarrow \ \mu, 
\notag \\
A_i \ &\longleftrightarrow \ \mu \mathcal{A}_i = \mu (- y \Omega, x \Omega),
\notag \\
H_z
\
&\longleftrightarrow
\
2 \mu \Omega.
\end{align}
This phenomenon is called the Barnett effect in nuclear matter \cite{Becattini:2020nfu, Becattini:2021lfq, Huang:2017pqe}.

\end{appendix}

\section*{Declaration of generative AI and AI-assisted technologies in the manuscript preparation process}
During the preparation of this work, the authors used ChatGPT
(OpenAI) in order to improve the English language, clarity, and
readability of the manuscript. After using this tool, the authors
reviewed and edited the content as needed and take full responsibility
for the content of the published article.

\bibliographystyle{elsarticle-num}
\bibliography{reference}

\end{document}